\documentclass[10pt,twocolumn,floatfix,prb,aps,showpacs,superscriptaddress,longbibliography]{revtex4-2}
\usepackage{graphicx,amsmath,amssymb,color, gensymb}
\usepackage{multirow}
\usepackage{nicefrac}
\usepackage[titletoc,title]{appendix}
\usepackage{amsmath}% for mathematical formulas
\usepackage{subfigure}
\usepackage{tabularx}
\usepackage{xcolor}
\usepackage{booktabs}
\usepackage{comment}
\usepackage[normalem]{ulem}
\usepackage[colorlinks,bookmarks=true,citecolor=blue,linkcolor=red,urlcolor=blue]{hyperref}
\usepackage{titlesec}
\usepackage[version=4]{mhchem}

\usepackage[T1]{fontenc}


\DeclareUnicodeCharacter{2215}{\slash{}}
\begin{document}
	
	\title{Exotic magnetism and persistent spin dynamics in a frustrated $J_\text{eff}=1/2$ triangular lattice antiferromagnet }
	
    \author{M. Barman}
    \affiliation{Department of Physics, Indian Institute of Technology Madras, Chennai, 600036, India}
     \author{K. Jakseti\v{c}}
    \affiliation{Jo\v{z}ef Stefan Institute, Jamova cesta 39, 1000 Ljubljana, Slovenia.}
     \affiliation{Faculty of Mathematics and Physics, University of Ljubljana, Jadranska ulica 19, 1000 Ljubljana, Slovenia.}
      \author{M. Pregelj}
       \affiliation{Jo\v{z}ef Stefan Institute, Jamova cesta 39, 1000 Ljubljana, Slovenia.}
    \affiliation{Faculty of Mathematics and Physics, University of Ljubljana, Jadranska ulica 19, 1000 Ljubljana, Slovenia.}
     \author{M. D. Le}
    \affiliation{ISIS facility, Rutherford Appleton Laboratory, Chilton, Didcot, OX11 0QX, Oxfordshire, UK.}
    \author{P. J. Baker}
    \affiliation{ISIS facility, Rutherford Appleton Laboratory, Chilton, Didcot, OX11 0QX, Oxfordshire, UK.}
     \author{A. Zorko}
    \affiliation{Jo\v{z}ef Stefan Institute, Jamova cesta 39, 1000 Ljubljana, Slovenia.}
     \affiliation{Faculty of Mathematics and Physics, University of Ljubljana, Jadranska ulica 19, 1000 Ljubljana, Slovenia.}
\author{P. Khuntia}
\email{pkhuntia@iitm.ac.in}
\affiliation{Department of Physics, Indian Institute of Technology Madras, Chennai, 600036, India}
\affiliation{Quantum Centre of Excellence for Diamond and Emergent Materials,
Indian Institute of Technology Madras, Chennai, 600036, India}

%\date{\today}
%TC:ignore
\begin{abstract} 
The delicate interplay between competing degrees of freedom, anisotropy, and frustration-induced strong quantum fluctuations in pseudospin-$J_\text{eff}=1/2$ rare-earth triangular lattice antiferromagnets offers a promising platform for the experimental realization of exotic states with nontrivial low-energy excitations. Here, we present thermodynamic, inelastic neutron scattering (INS), and muon spin relaxation ($\mu$SR) investigations of a frustrated magnet K$_{3}$NdTe$_{2}$O$_{9}$, in which Nd$^{3+}$ ions constitute a structurally perfect triangular lattice with no detectable site disorder. The experiments reveal the realization of a Kramers doublet ground state with $J_\text{eff}=1/2$ moments, well separated from the first excited state, which interact antiferromagnetically, with an exchange interaction of $\sim$ 0.6 K between the Nd$^{3+}$ moments in the triangular plane. The absence of oscillations and the so-called 1/3 plateau in the zero-field $\mu$SR asymmetry down to 50 mK rules out long-range magnetic ordering and spin freezing on the $\mu$SR time scale, respectively. The temperature dependence of zero-field $\mu$SR relaxation rate is well described by the Orbach relaxation mechanism, indicating the existence of fluctuating moments in the ground state of this frustrated magnet. Our results demonstrate exotic magnetism and persistent spin dynamics down to 50 mK. These observations establish this new family of frustrated rare-earth triangular-lattice antiferromagnets as a promising venue for the experimental realization of nontrivial quantum states with exotic low-energy excitations.

\end{abstract}
%TC:endignore
\maketitle

Correlated quantum materials are promising contenders to host emergent many-body phenomena driven by the competing interactions, external perturbations, topologically nontrivial band structure, reduced dimensionality and enhanced quantum fluctuations. Magnetic frustration in triangular, kagome or pyrochlore motifs arises from the incompatibility to simultaneously minimize all pairwise exchange interactions and leads to a highly degenerate manifold of low-energy ground states, such as quantum spin liquid (QSL)~\cite{khatua2023experimental,balents2010spin,lacroix2011introduction}. QSL is a highly entangled magnetic ground state characterized by the absence of any conventional symmetry-breaking phase transition down to absolute zero temperature ascribed to frustration induced strong quantum fluctuations~\cite{yan2011spin,khuntia2020gapless,PhysRevLett.116.107203,savary2017quantum,broholm2020quantum}. These states maintain persistent spin dynamics and host fractionalized excitations like spinons and Majorana fermions, exotic particles that move coherently within the emergent gauge fields~\cite{balents2010spin,khatua2023experimental}. Anderson initially proposed that the spin-$1/2$ Heisenberg antiferromagnet on a triangular lattice with nearest-neighbor interaction could support a QSL ground state within the resonating valence bond framework~\cite{anderson1973resonating}. Subsequent theoretical studies demonstrated that the isotropic spin-1/2 triangular-lattice Heisenberg antiferromagnet stabilizes a noncollinear $120^\circ$ magnetic order as its ground state, despite the strong quantum fluctuations present in the spin-1/2 limit~\cite{PhysRevLett.60.2531,PhysRevB.40.2727,PhysRevLett.69.2590,PhysRevLett.71.1629,PhysRevLett.99.127004}. Nevertheless, the introduction of additional interactions, such as next-nearest-neighbor exchange interactions~\cite{PhysRevB.93.144411,PhysRevB.92.041105,PhysRevB.92.140403}, anisotropic exchange couplings~\cite{PhysRevB.60.1064}, and magnetic anisotropy~\cite{PhysRevLett.112.127203,PhysRevB.72.045105,PhysRevB.95.165110}, can destabilize long-range magnetic order, thereby favoring quantum-disordered states including QSLs.

In this context, triangular-lattice antiferromagnets with next-nearest-neighbor exchange interactions or magnetic anisotropy provide one of the simplest archetypal systems for the experimental realization of exotic quantum states. In particular, $S = 1/2$ triangular-lattice antiferromagnets are prominent candidates for realizing QSL states, where frustration-induced strong quantum fluctuations suppress long-range magnetic ordering. Recently, rare-earth–based frustrated $4f$ triangular-lattice materials were found to offer an alternative route in which strong spin–orbit coupling and crystal electric field (CEF) generate highly anisotropic exchange interactions and effective pseudospin $J_\text{eff}=1/2$ anisotropies of the Ising, XY, or XYZ type~\cite{PhysRevB.51.8904,skomski2009anisotropy,xie2024rare,PhysRevB.92.041105,PhysRevLett.112.127203,arh2022ising,PhysRevB.94.035107,PhysRevB.94.201114,PhysRevB.95.165110,PhysRevLett.120.207203,10.21468/SciPostPhys.4.1.003,PhysRevLett.134.196702}. The strongly localized nature of the $4f$ orbitals result in weak exchange couplings in these frustrated magnets, enabling the stabilization of exotic magnetic phases that are highly tunable by external perturbations and thereby giving rise to rich magnetic phase diagram. These materials provide a promising venue for exploring unconventional thermodynamic responses highlighting the technological relevance~\cite{jaksetivc2026melting}, persistent spin dynamics within magnetically ordered states, field-induced quantum criticality, and the emergence of QSLs~\cite{PhysRevB.86.140405,PhysRevB.106.104408,PhysRevB.93.140408,PhysRevB.109.024427,1fqr-7njx,PhysRevB.111.155148,c3hd-pwby,PhysRevB.110.134401,PhysRevB.77.104413}.

In this vein, frustrated triangular-lattice antiferromagnets hosting Kramers rare-earth ions such as Nd$^{3+}$, Ce$^{3+}$, Yb$^{3+}$, etc. with an odd number of $4f$ electrons, are of particular interest because they host time-reversal-protected Kramers doublets that effectively realize pseudospin-$J_\text{eff}=1/2$ moments at low temperatures~\cite{kramers1930theorie,orbach1961spin}. Frustrated triangular-lattice families, such as the chalcogenides ARECh$_2$ ($A$ = alkali ion, RE = rare earth, Ch = chalcogen)~\cite{liu2018rare,PhysRevMaterials.3.114413,zhu2023fluctuating,bordelon2019field}, and the RETa$_7$O$_{19}$ family~\cite{arh2022ising,sibav2025optimized} have attracted considerable attention due to their near-ideal triangular lattices, strong magnetic anisotropy, and unconventional magnetic ground states. Recently, Nd-based triangular-lattice magnets have become an important platform to study frustration-driven quantum magnetism. For instance, Nd-based frustrated triangular lattices~\cite{arh2022ising,PhysRevB.111.155148,lv2025synthesis,xing2024candidate,gaudet2026vanishing} have attracted considerable attention due to signatures of persistent spin dynamics, field-induced magnetic states, strong Ising anisotropy, and quantum-disordered ground states with non-trivial low-energy excitations. A comprehensive understanding of the microscopic origin of exotic quantum phases, their associated excitations, and the role of external perturbations in frustrated rare-earth-based magnets remains relatively less explored. Therefore, the discovery and investigation of structurally perfect Kramers-ion-based $4f$ frustrated magnets attract significant attention for gaining deeper insight into the roles of competing degrees of freedom and anisotropic exchange interactions in driving emergent quantum phenomena. 

Here, we present synthesis, magnetization, specific heat, $\mu$SR, and INS studies of a structurally perfect frustrated triangular lattice antiferromagnet K$_{3}$NdTe$_{2}$O$_{9}$(henceforth KNTO). Magnetization measurements reveal the realization of a $J_\text{eff}=1/2$ Kramers doublet at low temperatures. Magnetization results indicates the presence of weak intraplane antiferromagnetic exchange interactions, $J_\text{ex}/k_\text{B}\sim 0.6(2)$ K, between the $J_\text{eff}=1/2$ moments, along with weak dipolar interactions, $J_\text{d}/k_\text{B} \sim$ 0.02 K. INS results along with thermodynamic data conform an anisotropic magnetic ground state doublet with a weak anisotropy, $\Delta g/g \sim 16\%$. Zero-field specific heat shows an anomaly at 81 mK where, however, the entropy release is only $\sim 8\%$ of $R\ln2$, which indicates a spurious phase or a fraction of spins undergo a static ordering, while the majority of spins maintain a dynamic state. $\mu$SR experiments reveal neither oscillations nor a 1/3 recovery in the asymmetry spectra, ruling out long-range magnetic order and spin freezing, down to 50 mK. Instead, this frustrated magnets exhibit persistent spin dynamics down to the lowest measured temperature. Such an unconventional magnetic ground state is driven by the geometric frustration and interplay between spin-orbit coupling and CEF.

 \begin{figure*}[t]
		\begin{center}
			\includegraphics[height=405pt, width=515pt]{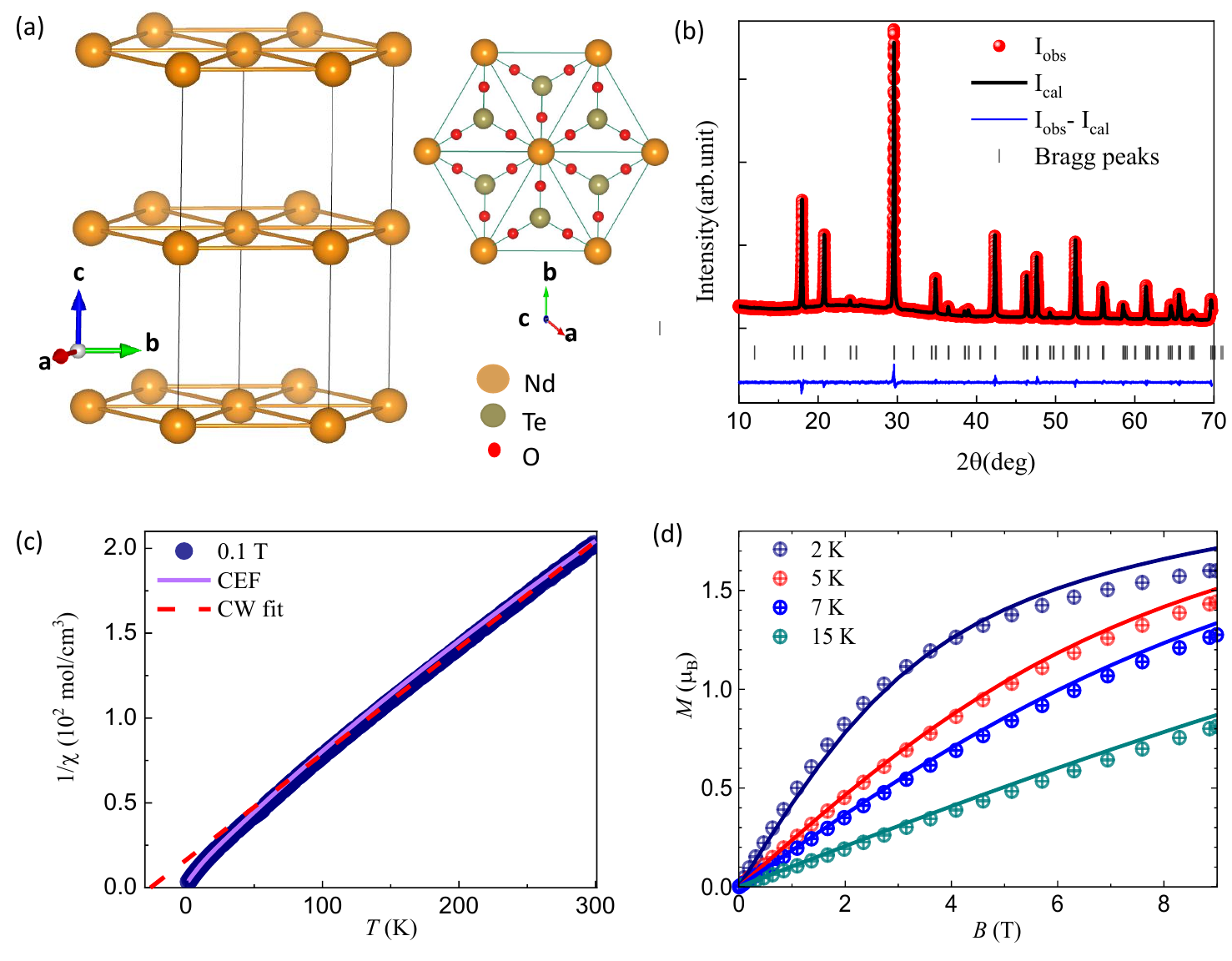}
			\caption{(a) Schematic of the triangular lattice in K$_{3}$NdTe$_{2}$O$_{9}$, where the magnetic Nd$^{3+}$ ions form a triangular network in the $ab$-plane with $d_\text{intra}=6.03$  $\text{\AA}$ and $d_\text{inter}=7.4$ $\text{\AA}$. The dominant magnetic superexchange interaction is mediated via Nd--O--Te--O--Nd exchange pathway. The right panel shows the triangular lattice formed by Nd$^{3+}$  moments in the crystal structure of KNTO, illustrating the geometric frustration inherent to the lattice with possible exchange pathways. (b) Rietveld refinement of the powder X-ray diffraction pattern of KNTO. (c)Temperature-dependent inverse susceptibility measured under an applied field of 0.1 T. The inverse susceptibility is fitted with the Curie–Weiss law at temperatures 100~K $\leq T \leq$ 300~K (dashed red line) and CEF fit(solid line). (d) The magnetization isotherms at several temperatures, and the solid lines correspond to the CEF fits.}
			\label{fig1}
		\end{center}
	\end{figure*}

   \begin{figure*}[t]
		\begin{center}
			\includegraphics[height=356pt, width=515pt]{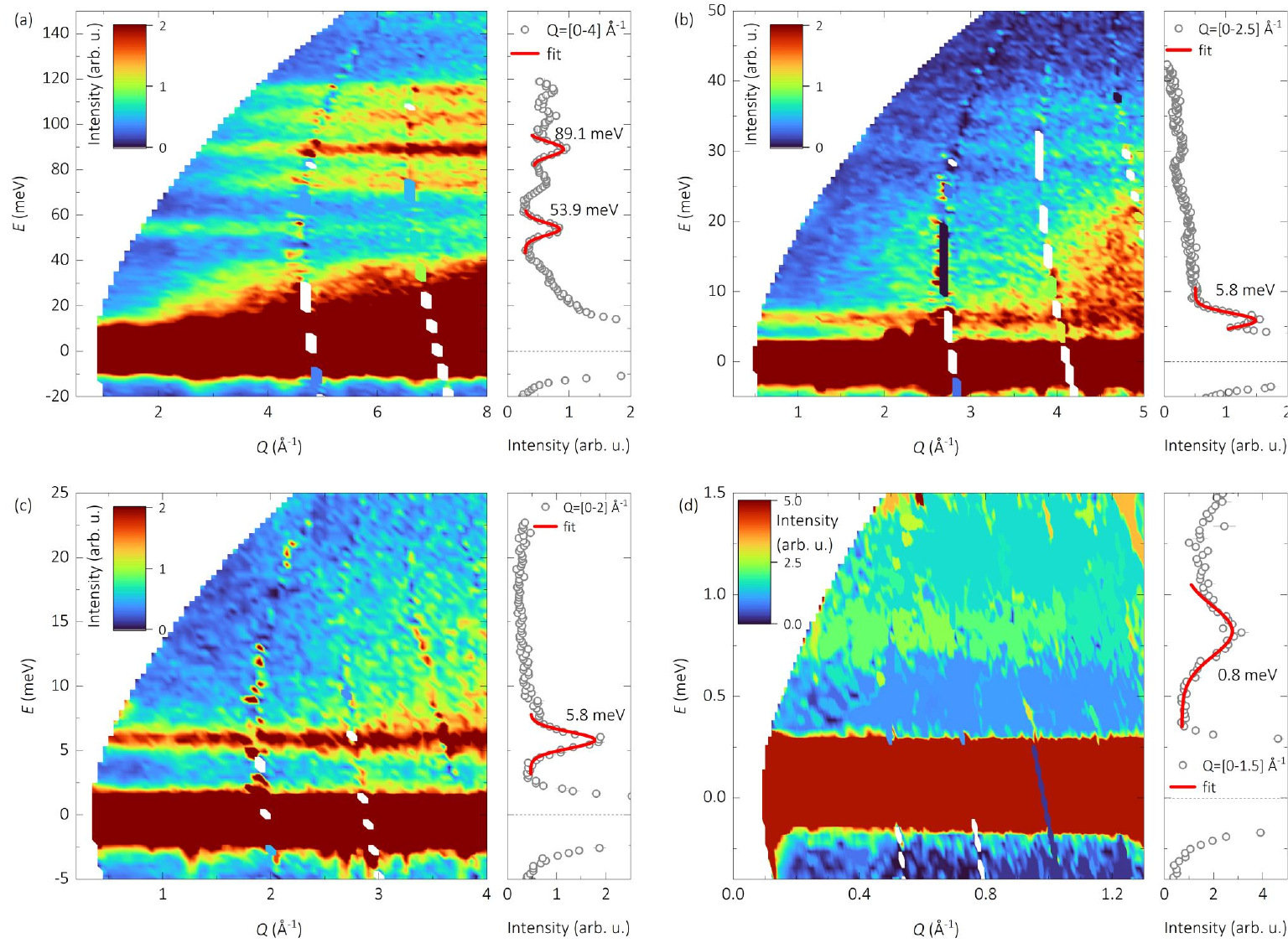}
			\caption{Inelastic neutron scattering (INS) intensity at 5K for the incident neutron energies of 180, 60, 29.7 and 2 meV. Besides the broad diffuse phonon excitations at larger scattering vectors Q, there are clear additional flat bands at low Q that correspond to CEF levels. Furthermore, the flat bands at E $>$ 60 meV which have maximal intensity at $\sim$7 meV, most likely correspond to molecular vibrations from adsorbed water. For clarity, we also show the energy transfer integrated across the low Q range to emphasize the flat bands. The red lines are Gaussian fits, performed to precisely determined the energies of the CEF excitations.}
			\label{fig2}
		\end{center}
	\end{figure*}

The polycrystal sample was synthesized by the solid-state reaction method (see SI details~\cite{SM}). The Rietveld refinement (Fig. \ref{fig1}(b)) of the XRD pattern reveals that KNTO crystallizes in the centrosymmetric hexagonal space group P63/mmc (No. 194) with lattice parameters $a=b=6.033(9)$ $\text{\AA}$; $c=14.786(5)$  $\text{\AA}$, without detectable atomic site disorder and the absence of any detectable impurity phases. The crystal structure in one unit cell is shown in Fig. \ref{fig1}(a), where the magnetic Nd$^{3+}$ ions are connected in a triangular lattice with the nearest-neighbor distance 6.03  $\text{\AA}$. The interlayer distance between two triangular layers is 7.4 $\text{\AA}$. The dominant magnetic superexchange interaction is likely mediated via Nd--O--Te--O--Nd path (right panel in Fig. \ref{fig1}(a)). The exchange interaction between Nd$^{3+}$ moments is expected to be weak owing to the longer exchange route, the presence of additional intermediate ions, and the highly localized character of the $4f$ orbitals (see SI \cite{SM} for more details). 

\begin{figure*}[t]
		\begin{center}
			\includegraphics[height=170pt, width=510pt]{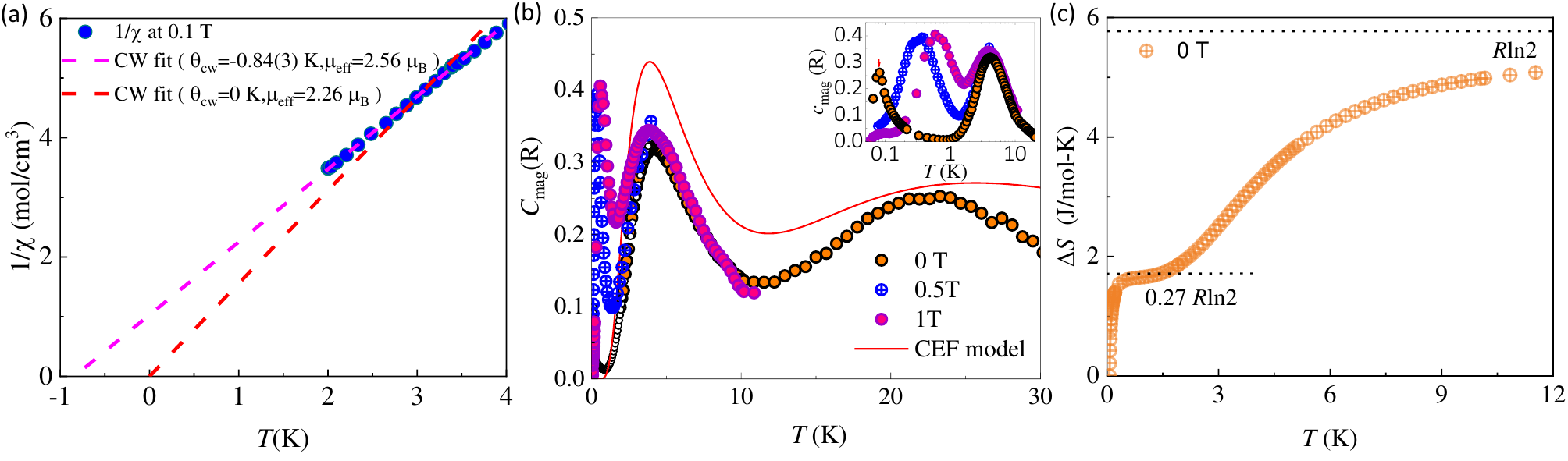}
			\caption{(a) Temperature-dependent inverse magnetic susceptibility recorded at an applied field of 0.1 T is fitted using the Curie–Weiss law (dashed magenta line) and non-interacting Curie–Weiss law with  $\Theta_\text{CW}=$ 0~K (dashed red line). (b)  Magnetic specific heat, obtained after subtracting the lattice contribution following the Debye model, measured under various magnetic fields. The solid red line is the CEF model. The inset highlights the low-temperature evolution of $C_{\mathrm{mag}}$ under applied magnetic fields. Zero-field magnetic specific heat exhibits a weak anomaly at 81 mK.  (c) The temperature dependence of magnetic entropy in zero field. It reaches 88$\%$ of the expected $R\ln2$ at 12 K for an effective $J_\text{eff}=1/2$ ground-state doublet. }
			\label{fig3}
		\end{center}
	\end{figure*}

The temperature dependence of the dc magnetic susceptibility data~(Fig. \ref{fig1}(c) and SI \cite{SM}) measured under an external magnetic field of 0.1 T shows no evidence of long-range magnetic ordering down to 1.9 K. The temperature dependence of the inverse magnetic susceptibility follows Curie–Weiss behavior above 150 K, which can be ascribed to CEF effects, while a change in slope below 75 K implies the emergence of another energy scale associated with magnetic exchange interactions at low temperatures. At high temperatures ($T > 150$~K), the susceptibility is dominated by thermally populated excited CEF levels. The Curie-Weiss fit of the inverse susceptibility data above 150 K yields an effective moment is $\mu_\text{eff}=3.57 ~\mu_\text{B}$, which is close to the free ion value for Nd$^{3+}$ effective moment $\mu_\text{eff}^\text{free}=3.62 ~\mu_\text{B}$. The Nd$^{3+}$ ion (4$f^3$; 4$I_{9/2}$; $L=6$; $S=$$\frac{3}{2}$; $J=$$\frac{9}{2}$) with an odd number of electrons and $J=9/2$ splits into five Kramers doublets by the CEF (See fig. \ref{fig2} and sTable III~\cite{SM}). Figure \ref{fig3}(a) shows the low temperature inverse susceptibility, and the Curie Weiss fit in the temperature range 1.9~K $\leq T \leq$ 3.8~K (see SI~\cite{SM,arh2022ising}), indicating that the susceptibility corresponds to the CEF ground-state doublet. The fit yields a Curie-Weiss temperature of $\Theta_\text{CW}=-0.84(3)$ K, and effective moment $\mu_\text{eff}=2.56$ $\mu_\text{B}$ corresponding to the powder average Lande, $g$ factor of $g=2.9$ at low temperatures. The CW temperature $\theta_{\mathrm{CW}}=-0.84$~K indicates the presence of weak antiferromagnetic exchange interactions of $J_\text{ex}/k_\text{B}=0.56(2)$~K between the nearest neighbor $J_\text{eff}=1/2$ moments in the $ab$-plane as estimated following the mean field approximation. The reduced effective moment at low temperatures indicates that only a subset of the CEF levels contributes to the magnetic response below 10 K. Higher-lying crystal field excitations are thermally depopulated at low temperatures, resulting in low-energy ground states dominated by the $J_\text{eff}$=1/2 degrees of freedom. This behavior suggests that the lowest $J_\text{eff}$=1/2 Kramers doublet ground state is well separated from the excited state. The absence of any bifurcation between the zero-field-cooled and field-cooled magnetization curves (Fig. S3(a)~\cite{SM}) rules out spin freezing down to 1.9 K, which is also confirmed by the absence of hysteresis in the magnetization isotherm data measured at 2 K. The magnetization isotherms measured at $T \ge 5$ K are well reproduced by a noninteracting single-ion CEF model (See SI \cite{SM}), however, the deviations observed at lower temperatures suggest the development of magnetic correlations, consistent with the magnetic susceptibility results.
It is worth mentioning that a pure non-interacting Curie model with $\Theta_{\mathrm{CW}}=0$ does not fit the low-temperature susceptibility well (Fig.~\ref{fig3}(a)), further suggesting that there are finite residual exchange interactions beyond the single-ion CEF description. The dipolar interaction in $ab$ plane is $J_\text{D}\sim \frac{\mu_0}{4\pi}\frac{\mu_\text{eff}^2}{k_\text{B}d_\text{nn}^3}= 0.018(6)$ K, here, $\mu_0$ is the vacuum permeability, $\mu_{\mathrm{eff}}$ is the effective magnetic moment at low temperatures, $k_{\mathrm{B}}$ is the Boltzmann constant, and $d_{\mathrm{nn}}$ denotes the nearest-neighbor distance between magnetic moments. The dipolar interaction is more than an order of magnitude smaller than the exchange interaction, indicating that exchange interactions play the dominant role in governing the low-temperature physics of this frustrated magnet.

The CEF level scheme determined from INS measurements provides additional insight into the magnetic properties of KNTO.  The dynamical structure factor $S(E,Q)$ (Fig.~\ref{fig2}), measured at 5 K, directly detects the CEF transitions. The INS intensity close to $E = 0$ is due to elastic scattering, while the scattering at $E > 0$, which increases with increasing scattering vector $Q$, is predominantly due to phonon excitations, as the magnetic correlations develop only at much lower temperatures owing to the strongly hybridized nature of the $4f$ orbitals typically observed in rare-earth-based magnets~\cite{arh2022ising}. The sharp flat bands at 0.8, 5.8, and 53.9 meV correspond to the CEF levels, as their intensity
decreases with increasing Q due to decreasing magnetic form factor $F^2(Q)$ of Nd$^{3+}$ ions. The additional flat bands at $\sim$ 75, 105, 90 and 115 meV, the intensity of which increases with increasing Q and exhibits a maximum at $\sim$~7 $\text{\AA}^{-1}$, that most likely correspond to molecular vibrations from adsorbed water~\cite{hall1981inelastic,miskowiec2019analysis}. Nevertheless, since Nd$^{3+}$ ion is expected to host four excited CEF doublets, the significant intensity of the flat band at 89.1 meV at the lowest accessible $Q$, may arise from the overlapping CEF excitations. In order to determine these parameters, allowed by the dihedral D$_\text{3d}$ point symmetry at the Nd$^{3+}$ in KNTO, we simultaneously fitted the observed CEF transition energies determined from the INS spectra (Fig.~\ref{fig2}), magnetic susceptibility, magnetization and specific heat data. The fit reproduces all experimental data well (see SI~\cite{SM} for more details).

The CEF analysis reveals the presence of five Kramers doublets, as expected for $J=9/2$ multiplet of Nd$^{3+}$ (SI Table II~\cite{SM}). All four excited doublets from the CEF ground state to the excited states occur for energy transfers below 100 meV. While the lowest three excited doublets are obvious from the INS data, the highest-energy excitation appears to be overlapping with the vibrational mode of the adsorbed water. The composition of the CEF ground-state wave function gives rise to a moderate $g$ factor anisotropy ($\Delta g/g \sim 16\%$) (SI Table-II~\cite{SM}). 
  \begin{figure*}[t]
		\begin{center}
			\includegraphics[height=407pt, width=515pt]{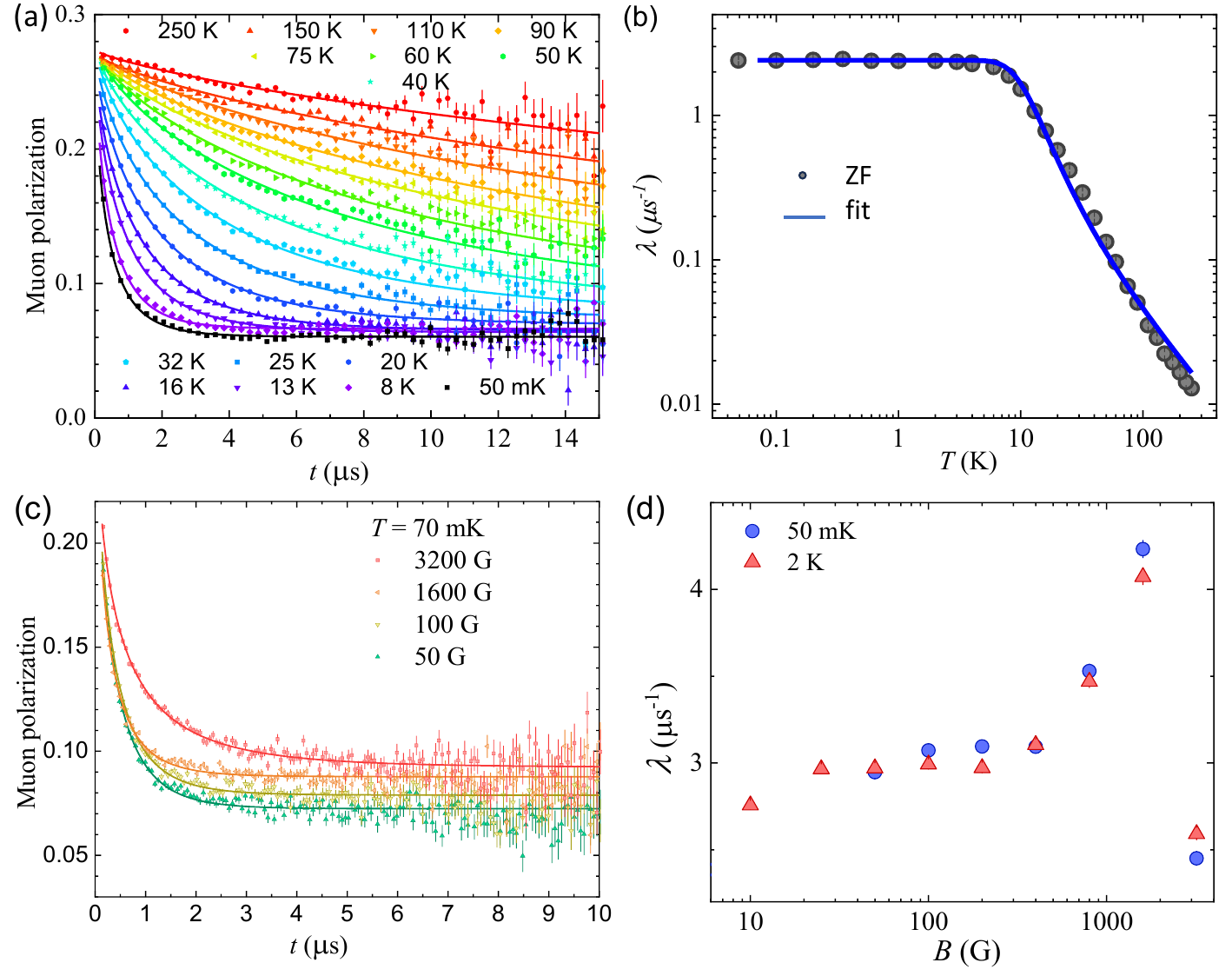}
			\caption{(a) Zero-field $\mu$SR asymmetry spectra at selected temperatures, demonstrating the temperature evolution of the muon spin polarization. The solid lines are fits to a stretched exponential relaxation function, as discussed in the text (SI \cite{SM}). (b) The temperature dependence of $\mu$SR rate $\lambda$ in zero field. The solid blue line shows the expected thermally activated behavior of the crystal electric field excitations as discussed in the text. (c) $\mu$SR asymmetry spectra in different applied longitudinal magnetic fields at 70 mK, demonstrating partial decoupling of muon relaxation with field. The solid lines are fits to a stretched-exponential relaxation function, $P_\text{LF}(t) = \exp[ - (\lambda_\text{LF} t)^{\beta_\text{LF}}$ where $\lambda_{\mathrm{LF}}$ and $\beta_{\mathrm{LF}}$ are the muon relaxation rate and stretching exponent, respectively. (d) Relaxation rate as a function of the longitudinal field at $50$ mK and $2$ K.}
			\label{fig4}
		\end{center}
	\end{figure*}

To investigate the ground state, low-energy excitations and spin correlations in KNTO, we performed the specific heat measurement down to 67 mK. The magnetic specific heat $C_\text{mag}$ is shown in Fig. \ref{fig3}(b) after subtracting the lattice contribution estimated using the Dybye model, due to the unavailability of a non-magnetic analogue of KNTO. The magnetic specific heat shows a weak $\lambda-$ type anomaly at 81 mK in zero field (inset of Fig. \ref{fig3}(b)). The associated entropy release is only $\sim8\%$ of $R\ln2$,  indicating that a fraction of moments possibly becomes static below the transition while the majority of the Nd$^{3+}$ moments maintain a fluctuating state down to 67 mK. The frustration parameter reveals strong degree of frustration, $f=\frac{|\theta_\text{CW}|}{T_\text{N}}$ $\sim$ 10, as suitable for a strongly frustrated triangular lattice. Upon application of a magnetic field, a broad peak appears at $\sim$0.5 K at 0.5 T, and progressively broadens and shifts to higher temperatures with increasing field. This scenario is consistent with a Schottky-type contribution arising from the Zeeman splitting of the ground-state Kramers doublet. In addition to the low-temperature anomaly in $C_\text{mag}$,  two broad maxima are observed, one around 4 K and another near 22 K, which are attributed to the thermal population of excited CEF levels. As shown in Fig. \ref{fig3}(c), the magnetic entropy reaches 5.1 J/mol-K at high temperatures in zero field,  corresponding to 88$\%$ of the expected $R\ln2$ for $J_\text{eff}$=1/2 moment. One plausible scenario for the reduced magnetic entropy arise from residual short-range spin correlations and or uncertainties in the lattice subtraction owing to the unavailability of a non-magnetic analogue. Nevertheless, the temperature evolution of entropy in zero field is consistent with a ground state that is governed by the lowest Kramers doublet with $J_\text{eff}$ = 1/2. Another plateau-like feature is observed in the zero-field magnetic entropy around 350 mK, with only 27$\%$ (1.6 J/mol-K) of the $R\ln2$ entropy released up to 1.7 K. Above this temperature, the entropy increases gradually. Such behavior has been reported in several frustrated magnets, including rare-earth-based intermetallic compound CeAlSi~\cite{zxc4-6xvk} and the triangular-lattice antiferromagnet NiGa$_2$S$_4$~\cite{nakatsuji2005spin}. One possible scenario that reconciles these features in the magnetic entropy is the presence of a highly degenerate low-energy manifold in these frustrated magnets.

To shed further light on the magnetic ground state and associated low-energy excitations, we carried out $\mu$SR measurements on powder sample down to 50 mK. The muon asymmetry exhibits a monotonic relaxation without any oscillatory behavior (Fig. \ref{fig4}(a)), indicating the existence of fluctuating local magnetic field due to electronic moments at the muon stopping sites.~\cite{le2011muon}. Frustrated magnets in a magnetically long-range ordered state commonly exhibit characteristic oscillations of the muon asymmetry owing to the presence of well-defined static internal magnetic fields. Although the specific-heat data suggest the appearance of an anomaly at 81~mK, no oscillatory signal is detected in the $\mu$SR spectra down to 50~mK. Furthermore, the absence of the so-called 1/3 tail at long times in the muon asymmetry rules out static local fields on $\mu$SR time scale down to 50 mK.

The ZF $\mu$SR relaxation rate $\lambda$ is nearly temperature independent below $\sim 9$~K (Fig. \ref{fig4}(b)), indicating that the spin dynamics is dominated by the CEF ground-state Kramers doublet. Upon further increase of the temperature, $\lambda$ rapidly decreases, indicating a crossover to a thermally activated relaxation regime. Such behavior is typically observed in rare-earth magnets, where low-temperature spin fluctuations are driven by the CEF ground-state Kramers doublet, while the higher-temperature dynamics are mediated by thermally populated CEF excited levels. The ZF $\mu$SR relaxation process can be described within an Orbach-activated relaxation process where the rate of relaxation is given by~\cite{orbach1961spin,arh2022ising}:$\frac{1}{\lambda}=\frac{1}{\lambda_0}+\frac{\eta}{\mathrm{exp}(\Delta_\mathrm{\mu SR}/T)-1}$, where $\lambda_{0}$ is the temperature independent contribution to the relaxation in the ground-state Kramers doublet, $\eta$ represents the amplitude parameter of relaxation governed by the Orbach process, $\Delta_\mathrm{\mu SR}$ corresponds to the CEF gap and $k_{\mathrm{B}}$ is the Boltzmann constant. The fit yields $\lambda_{0}=2.4~\mu$s$^{-1}$, $\eta=10.57~\mu$s , and $\Delta_\mathrm{\mu SR}=39$ K. The extracted activation energy $\Delta_\mathrm{\mu SR}=39$ K, which is a bit larger than that obtained from INS.  Nevertheless, it indicates that the ground-state Kramers doublet is well separated from the excited CEF states, consistent with several other rare-earth ion-based frustrated magnets~\cite{arh2022ising,PhysRevB.111.155148}.

Longitudinal-field (LF) $\mu$SR measurements were performed at 70~mK to probe the dynamic component of the muon spin relaxation. In the presence of an applied LF, the muon polarization is dominated by fluctuations of the electronic $J_{\mathrm{eff}}=1/2$ moments coupled to the implanted muons. As shown in Fig.~\ref{fig4}(c), the muon polarization is not completely recovered even at the highest LF $= 0.32$~T, reflecting that the relaxation is not completely decoupled by the applied magnetic field. This scenario is consistent with the presence of persistent spin dynamics as opposed to static internal fields. The field dependence of the relaxation rate measured at $50$ mK and $2$ K is shown in Fig. \ref{fig4}(d). The relaxation rate is nearly constant in low fields, and increases slowly with the field, which suggests slowing down of spin fluctuations.

In summary, we have investigated magnetic properties of the disorder-free geometrically frustrated triangular-lattice antiferromagnet K$_3$NdTe$_2$O$_9$ using magnetization, specific heat, $\mu$SR and INS measurements. Our results establish that the ground-state Kramers doublet is well separated from the excited CEF levels, indicating that the low-energy magnetic properties are governed by the weakly interacting effective  $J_\text{eff}=1/2$ moments. The INS experiments reveal a weak $g$-factor anisotropy in this frustrated triangular lattice antiferromagnet. The specific-heat measurements detect a nontrivial ground state with persistent low-energy excitations, consistent with a highly degenerate ground-state manifold. The $\mu$SR measurements unveil persistent dynamical state down to 50 mK, confirming the absence of static long-range magnetic order or spin freezing on the $\mu$SR time scale. Taken together, these observations highlight the intricate interplay of geometric frustration, competing exchange interactions, strong spin-orbit coupling, CEF, and weak anisotropy in stabilizing exotic magnetism and persistent low-energy spin dynamics in the broad family of rare-earth-based frustrated antiferromagnets and provide a promising venue for testing and developing advanced theoretical models.

P.K. acknowledges funding from the Science and Engineering Research Board and Department of Science and Technology, India through research grants. We acknowledge the financial support of the Slovenian Research and Innovation Agency through the Programs No. P1-0125 and Projects No. N1-0148 and J1-50008.

\bibliography{KNTObib}

\end{document}